\documentclass[aps,preprintnumbers,prl,amsmath,amssymb,twocolumn,superscriptaddress]{revtex4-1}

\usepackage{graphicx}
\usepackage{dcolumn}
\usepackage{bm}
\usepackage{color}
\usepackage{amssymb}   
\usepackage{amsthm}

\usepackage[normalem]{ulem}

\usepackage{hyperref}
\hypersetup{colorlinks = true,citecolor = blue,urlcolor = black}

\begin{document}

\title{Length-scales in sheared soft matter depend sensitively on molecular interactions}

\author{A. Scacchi}
\affiliation{Interdisciplinary Centre for Mathematical Modelling and Department of Mathematical Sciences, Loughborough University, Loughborough LE11 3TU, UK}
\affiliation{Department of Chemistry and Materials Science, Aalto University, P.O. Box 16100, FI-00076 Aalto, Finland}
\affiliation{Department of Applied Physics, Aalto University, P.O. Box 11000, FI-00076 Aalto, Finland}
\author{M. G. Mazza}
\affiliation{Interdisciplinary Centre for Mathematical Modelling and Department of Mathematical Sciences, Loughborough University, Loughborough LE11 3TU, UK}
\affiliation{Max Planck Institute for Dynamics and Self-Organization, Am Fa{\ss}berg 17, 37077 G\"ottingen, Germany}
\author{A. J. Archer}
\affiliation{Interdisciplinary Centre for Mathematical Modelling and Department of Mathematical Sciences, Loughborough University, Loughborough LE11 3TU, UK}

\begin{abstract}
The structure and degree of order in soft matter and other materials is intimately connected to the nature of the interactions between the particles. One important research goal is to find suitable control mechanisms, to enhance or suppress different structures. Using dynamical density functional theory, we investigate the interplay between external shear and the characteristic length-scales in the interparticle correlations of a model system. We show that shear can controllably change the characteristic length-scale from one to another quite distinct value. Moreover, with specific small changes in the form of the particle interactions, the applied shear can either selectively enhance or suppress the different characteristic wavelengths of the system, thus showing how to tune these. Our results suggest that the nonlinear response to flow can be harnessed to design novel actively responsive materials.
\end{abstract}

\maketitle

The manipulation and control of matter lies at the heart of much physical and technological progress~\cite{ruokolainen1998switching}. Soft matter is particularly captivating because of its high responsiveness to external stimuli~\cite{kato2006functional}. The control of order in soft materials can lead to novel functionalities, associated with dynamical structural changes, including colloidal self-assembly, and by combining fluids and light, various photonic devices, e.g.\ adaptive optical lenses, have been made~\cite{muvsevivc2006two, whitesides2006origins, yamada2008photomobile, humar2009electrically, ravnik2011three, cipparrone2011chiral, sacanna2011shape, lagerwall2012new, monat2007integrated}. Developing materials with structures that can be precisely tuned depends on having control of the characteristic length-scales~\cite{schmidt2011photonic, taylor2013small,caleap2014acoustically}; the fluidic nature of soft matter can make them tunable and reconfigurable~\cite{monat2007integrated}.

A major challenge to control and manipulate nonequilibrium soft matter is ascertaining the link between the microscopic structure and the macroscopic flow behavior~\cite{leePRL2019, guNat2018}. When a complex fluid is subject to shear, novel structure can emerge. Recent studies include for lamellar phases~\cite{diat1995layering}, colloids~\cite{cohen2004shear,gerloff2017shear,khabaz2017shear,varga2019hydrodynamics}, colloidal gels~\cite{koumakis2015tuning,moghimi2017colloidal,massaroSM2020}, star polymers~\cite{ripoll2006star,liu2018shear}, biomolecules~\cite{schneider2007shear,singh2009fluid,shan2019shear,zhou2019effects}, biofluids~\cite{liu2019nanoparticle}, hydrogels~\cite{ke2019shear}, and graphene particles~\cite{hong2018shear,zhaoNat2020}. Neutron and X-ray scattering methods have proven valuable probes of the structure and rheology in soft matter~\cite{pozzo2008macroscopic,singh2009fluid,eberle2012flow,leheny2015rheo,westermeier2016connecting,narayananCrystRev2017,liu2018shear,gibaudPRX2020,andrianoJRheo2020,guNat2018}. Surprisingly, little is known about the stability under shear of such structures and how this depends on the particle interactions, though the question is of relevance e.g.\ for ink-jet print technologies~\cite{fuller2002ink,le1998progress} and biological microfluidics~\cite{domachuk2010bio}. To understand stability and other properties of sheared systems, a well-founded dynamical theory is necessary.

Here, we offer general arguments for a model complex fluid exhibiting at least two characteristic length-scales, which predict counterintuitive behavior. We employ dynamical density functional theory (DDFT) because it is an efficient approximate method to determine a system's dynamical response to external fields and external driving in terms of the particle number density $\rho(\mathbf{r},t)$, as a function of position $\mathbf{r}$ and time $t$, which is a quantity readily accessible to experiments. Thus, DDFT yields both the equilibrium and the nonequilibirum microscopic structure of the material, which includes information about characteristic length-scales of the system. The original formulation of DDFT, which can be derived either from the Langevin equation of motion~\cite{marconi_tarazona_1, marconi_tarazona_2} or from the Smoluchowski equation \citep{archer_evans, ArRa04}, considered time-dependent external potentials. Rauscher {\it et al.}\ generalized theory to the treatment of external flow fields (e.g.~shear), by incorporating an additional term describing the affine solvent flow field \citep{rauscher_penna_2007}. However, this form is incapable of describing interaction-induced currents orthogonal to the affine flow, crucial in the case of e.g.\ laning. Using a dynamical mean-field approximation, this issue has been successfully addressed in  \citep{kruger2011controlling, brader_kruger_2011, scacchi_kruger_brader_2016}, and applied to study the laning instability \citep{scacchi2017dynamical}. The DDFT equation for the time evolution of $\rho$ is
\begin{equation}
\frac{\partial \rho}{\partial t}+\nabla\cdot(\rho \textbf{v})=\Gamma\nabla\cdot\left(\rho\nabla\frac{\delta \mathcal{F[\rho]}}{\delta\rho}\right),\label{DDFT}
\end{equation}
where $\textbf{v}(\mathbf{r},t)$ is the solvent velocity, $\mathcal{F}$ the Helmholtz free energy functional from equilibrium density functional theory \cite{evans_79, evans_92, hansen_mcdonald} and $\Gamma=\beta D$ is the particle mobility, where $D$ is the diffusion coefficient and $\beta=(k_BT)^{-1}$; $k_B$ is Boltzmann's constant and $T$ is the temperature. The Helmholtz free energy is composed of three terms:
\begin{align}
  \mathcal{F}[\rho]=k_BT\int d\mathbf{r} \rho(\ln(\Lambda^n\rho)-1) 
+ \mathcal{F}^{exc}[\rho] +
  \int d\mathbf{r}\rho V_{ext}.
\label{helmholtz}  
\end{align}
The first term is the ideal-gas part, where $\Lambda$ is the thermal de Broglie wavelength and $n$ is the dimensionality of the system. $\mathcal{F}^{exc}[\rho]$ is the excess Helmholtz free energy functional, which incorporates the influence of the interparticle interactions, and $V_{ext}(\mathbf{r})$ is the external potential.

From Eq.~(\ref{helmholtz}), the functional derivative in Eq.~(\ref{DDFT}) is
\begin{equation}
\frac{\delta\mathcal{F}[\rho]}{\delta\rho}=k_\mathrm{B}T\ln[\Lambda^n\rho]-k_\mathrm{B}Tc^{(1)}+V_{ext},
\end{equation}
where $c^{(1)}(\mathbf{r})\equiv -\beta \frac{\delta\mathcal{F}^{exc}[\rho]}{\delta\rho}$
is the one-body direct correlation function~\citep{evans_79,evans_92, hansen_mcdonald}. We write the velocity field as $\mathbf{v}(\mathbf{r},t)=\mathbf{v}^{aff}(\mathbf{r},t)+\mathbf{v}^{fl}(\mathbf{r},t)$, which is a sum of the affine flow $\mathbf{v}^{aff}$ and a particle induced `fluctuation' flow, $\mathbf{v}^{fl}$, which incorporates the influence of the forces a pair of approaching particles with different velocities experience as they flow around each other. $\mathbf{v}^{fl}$ is not known exactly. However, we do know it must be a functional of the fluid density profile $\rho$ and it must be zero when the fluid density is uniform, $\rho(\mathbf{r},t)=\rho_b$, corresponding to an equilibrium homogeneous state with zero particle flux. Assuming that it is possible to make a functional Taylor expansion, the fluctuation term can be written as
\begin{equation}
\textbf{v}^{fl}[\rho(\mathbf{r})]=\textbf{v}^{fl}[\rho_b]+\int d\mathbf{r}' \delta\rho(\mathbf{r}') \frac{\delta \textbf{v}^{fl}[\rho(\mathbf{r})]}{\delta\rho(\mathbf{r}')}\bigg|_{\rho_b}+\mathcal{O}(\delta\rho^2),
\end{equation}
where $\delta\rho(\mathbf{r})=\rho(\mathbf{r})-\rho_b$. The first term is zero, and truncating after the second term we obtain the form suggested in Ref.~\cite{brader_kruger_2011}, i.e.
\begin{equation}
\textbf{v}^{fl}(\mathbf{r})=\int d\mathbf{r}'\delta\rho(\mathbf{r}')\boldsymbol{\kappa}(\mathbf{r}-\mathbf{r}'),\label{non-affine}
\end{equation}
where $\boldsymbol{\kappa}(\mathbf{r}-\mathbf{r}')=\frac{\delta \textbf{v}^{fl}[\rho(\mathbf{r})]}{\delta\rho(\mathbf{r}')}\big|_{\rho_b}$ is the flow kernel. This quantity has been introduced relatively recently, but its identification has already led to a series of advances in the areas of nonlinear rheology and colloidal fluids~\citep{kruger2011controlling, brader_kruger_2011, scacchi_kruger_brader_2016, scacchi2017dynamical, scacchi2018flow, stopper2018nonequilibrium}. We venture to compare the current state-of-the-art knowledge of the flow kernel with the state of affairs when Ornstein an Zernike first identified the direct correlation function $c^{(2)}(r)$ in the early stages of equilibrium liquid state theory~\cite{hansen_mcdonald}: they knew it is important, but at first they knew very little about its full form. The kernel function $\boldsymbol{\kappa}(\mathbf{r}-\mathbf{r}')$ describes the influence on the velocity of particles colliding with their neighbors. In the bulk $\boldsymbol{\kappa}$ is translationally invariant, but in the presence of confinement, this is no longer the case, especially at distances of order of a particle diameter from any confining substrate. Physical interpretation of $\boldsymbol{\kappa}$ was given in~\cite{brader_kruger_2011}, where for a fluid of hard-spheres the authors obtained an expression for $\boldsymbol{\kappa}$ by means of a geometrical construction. More recently, in~\cite{scacchi_kruger_brader_2016} the authors derived an expression for the kernel function in an exactly solvable low density limit, again for hard-spheres. What can be inferred more generally is that the kernel function $\boldsymbol{\kappa}(\mathbf{r}-\mathbf{r}')$ is an odd function of its argument, with a range that is comparable to the interaction potential. 

Our focus here is on systems of soft penetrable particles interacting via pair potentials which model the effective interactions between polymeric macromolecules such as star polymers, dendrimers or micelle forming block copolymers in solution \cite{likos2001effective}. Note, however, that our findings can be generalized to all systems with multiple length-scales. Assuming a steady shear with flow in the $x$-direction and a linear gradient in the $y$-direction, one can calculate the dispersion relation $\omega$ perpendicular to the flow~\cite{scacchi2017dynamical} (see the Appendix for more details). $\omega$ determines the growth or decay of density modulations $\sim\exp(\omega t+i\mathbf{k}\cdot\mathbf{r})$, with wavevector $\mathbf{k}$. For simplicity and without loss of generality, we treat the system as being two dimensional, so $\mathbf{k}=(k_x,k_y)$ and $\omega$ can be written as 
\begin{equation}
\omega(0, k_y)=-\rho_b k_y \alpha_y(k_y)-\frac{k_y^2D}{S(k_y)},
\label{dispersion_relation_final}
\end{equation}
where $\alpha_y(k_y)=\int d\mathbf{r} \sin(k_y y)\kappa_y(\mathbf{r})$ and $S(k)$ is the bulk fluid static structure factor. When the shear rate $\dot{\gamma}=0$, then $ \alpha_y(k_y)=0$. Also, since $S(k)>0$ for all wavenumbers $k=|\mathbf{k}|$, Eq.\ (\ref{dispersion_relation_final}) shows that $\omega<0$ for all $k$ when $\alpha_y=0$. Of course, this means that when $\dot{\gamma}=0$ the uniform liquid is linearly stable, as expected. However, for $\dot{\gamma}>0$ this is not necessarily the case. In the case of a system with one length scale, a sufficiently large $\dot{\gamma}$ can result in $\omega>0$ for a small band of wave numbers at certain characterisitic wavenumbers, leading to a `laning instability' \cite{kruger2011controlling, brader_kruger_2011, scacchi_kruger_brader_2016, scacchi2017dynamical, scacchi2018flow, stopper2018nonequilibrium}. For systems with multiple intrinsic length-scales, the dispersion relation exhibits several distinct peaks, each corresponding to a different length-scale. For $\dot{\gamma}>0$ the effect of $\boldsymbol{\kappa}$, and thus of the function $\alpha_y(k_y)$, is to increase or decrease the height of the peaks in the dispersion relation. This shows it is possible to manipulate the length-scales inherent in the structure of the system and also change the length-scale for which the system is linearly unstable, and so change e.g.\ the spatial frequencies of the oscillations in the laned state. 

We illustrate these general observations for the family of fluids composed of particles interacting via the Barkan-Engel-Lifshitz (BEL) pair potential~\cite{barkan2014controlled}, which has multiple characteristic length-scales. This model system offers computational convenience and the possibility to study complex external shear. The multiple length-scales in the BEL potential can lead to the formation of multiple complex equilibrium structures, such as  quasicrystals. While in the following we specialize to results based on the BEL potential, our results can be generalized to a wide range of multiple length-scales systems. The BEL pair potential has the form
\begin{equation}\label{pairpotential}
\phi(r)=\varepsilon e^{-\frac{1}{2}\sigma^2 r^2}\left(c_0+c_2 r^2+c_4 r^4+c_6r^6+c_8 r^8\right),
\end{equation}
where the parameter $\varepsilon$ defines the overall strength of the repulsion and the other coefficients can be adjusted so that the structures formed exhibit two characteristic length-scales. An additional reason for choosing the BEL model is the fact that the excess Helmholtz free energy functional can be approximated with surprising accuracy by the simple mean-field form 
\begin{align}
  \mathcal{F}^{exc}[\rho]= \frac{1}{2}\int d\mathbf{r} \int d\mathbf{r}' \rho(\mathbf{r})\rho(\mathbf{r}')\phi(\mid\mathbf{r}-\mathbf{r}'\mid).
\label{meanfield}  
\end{align}
With this approximation, we have $S(k)^{-1}=1+\rho_b\beta\hat{\phi}(k)$ in Eq.~\eqref{dispersion_relation_final}, where $\hat{\phi}(k)$ is the Fourier transform of the pair potential \cite{scacchi2017dynamical}. We choose the simplest ansatz for the flow kernel $\boldsymbol{\kappa}(\mathbf{r})$ compatible with the symmetry of the system,
\begin{equation}
\boldsymbol{\kappa}(\mathbf{r})=-\beta f(\dot{\gamma},\rho)\nabla\phi(\mathbf{r}).\label{kernel_grad_phi}
\end{equation}
Recent results for hard-disks imply that the prefactor $f\sim\dot{\gamma}^{0.8}$ \cite{JS2020}. However, here we further simplify by assuming $f=a\dot{\gamma}$, where $a=1$ is constant \cite{scacchi2017dynamical}. 

\begin{figure}[t]
\includegraphics[width=\columnwidth]{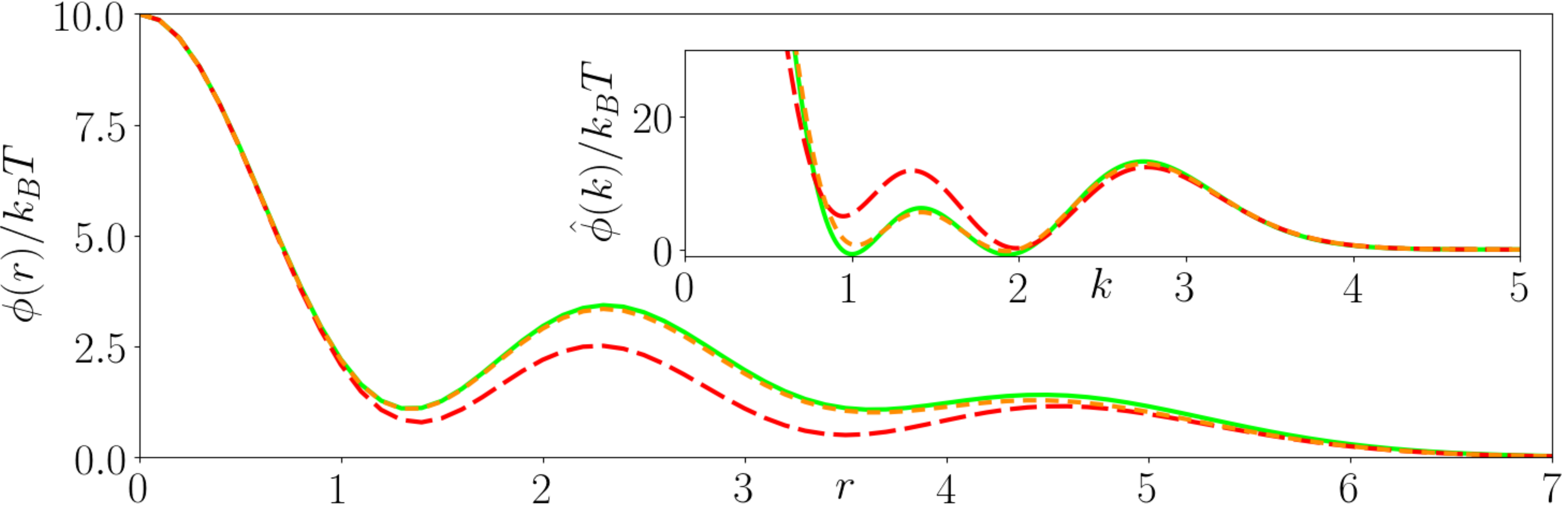}\caption{Three rather similar interparticle pair potentials with extremely different behavior under shear, cases 1-3 discussed in the text, which are respectively the solid (green), short-dashed (orange) and long-dashed (red) lines. The inset displays the Fourier transform of the pair potentials $\hat{\phi}(k)$.}\label{pairpot}
\end{figure}

We now show that small changes in the pair potential $\phi$, corresponding to small changes in the chemistry or molecular architecture of the polymeric macromolecules, can result in substantially different effects. We illustrate this fact by varying the set $\mathcal{C}\equiv\lbrace \sigma+\delta\sigma, c_0, c_2, c_4+\delta c_4, c_6, c_8 \rbrace$, where $\delta\sigma$ and $\delta c_4$ are the small variations considered and where the reference values $\lbrace \sigma, c_0, c_2, c_4, c_6, c_8 \rbrace$ are those proposed in \cite{barkan2014controlled}, i.e.\ $\mathcal{C}_{R} \equiv  \lbrace 0.771, 1, -1.095, 0.4397, -0.04927, 0.001831 \rbrace$, which lead to the system having two characteristic length-scales $2\pi/k_1$ and $2\pi/k_2$, where the corresponding wavenumbers are $k_1=1$ and $k_2=1.93\approx 2$. Specifically, we discuss three different cases, illustrated in Fig.~\ref{pairpot}. We emphasize that the three pair potentials appear structurally very similar, but give rise to three very different scenarios under shear: In \emph{Case 1}, we see below that the external shear causes the system to switch from one characteristic length-scale being dominant to the other. In \emph{Case 2}, both length-scales are amplified by shear and thus their prominence in the overall structure of the fluid is enhanced. In this case, if the shear rate is sufficiently large it is possible to form a laned state. In the last \emph{Case 3} shear has the \emph{opposite} effect, increasing the damping of both characteristic length-scales, making the uniform liquid even more stable under shear. Perturbations of the steady state are absorbed more rapidly in the nonequilibrium case. We present hereafter the equilibrium and steady-state (under shear) density profiles together with corresponding dispersion relations for different shear rates $\dot{\gamma}$. We denote as $\dot{\gamma}_{c}$ the threshold value of the shear rate at which the system becomes linearly unstable (for one of the length-scales). For all three cases the system is confined between a pair of parallel (soft) purely repulsive walls, exerting the external potential
\begin{equation}
V_{ext}(y)=
\begin{cases}
V_0 e^{-\frac{y^{3/2}}{4}}, &\mbox{$0<y<\frac{L}{2}$} 
\\ 
V_0 , &\mbox{$y<0$ or $y>L$}
\\
V_0 e^{-\frac{(y-L)^{3/2}}{4}}, &\mbox{$\frac{L}{2}<y<L$}

\end{cases},\label{external_potential}
\end{equation}
where the $y$-axis is perpendicular to the surfaces of the walls, $L=172$ is the distance between the walls and $V_0$ is the repulsion strength. The softness of $V_{ext}$ aids the numerical stability of the calculations. For the presented cases the values of $\beta V_0$ are chosen to be different only for visualization purposes and are specified below. We fix the bulk fluid density to be $\rho_b=0.5$ in all our calculations.

\begin{figure*}[t]
\includegraphics[width=1.\linewidth]{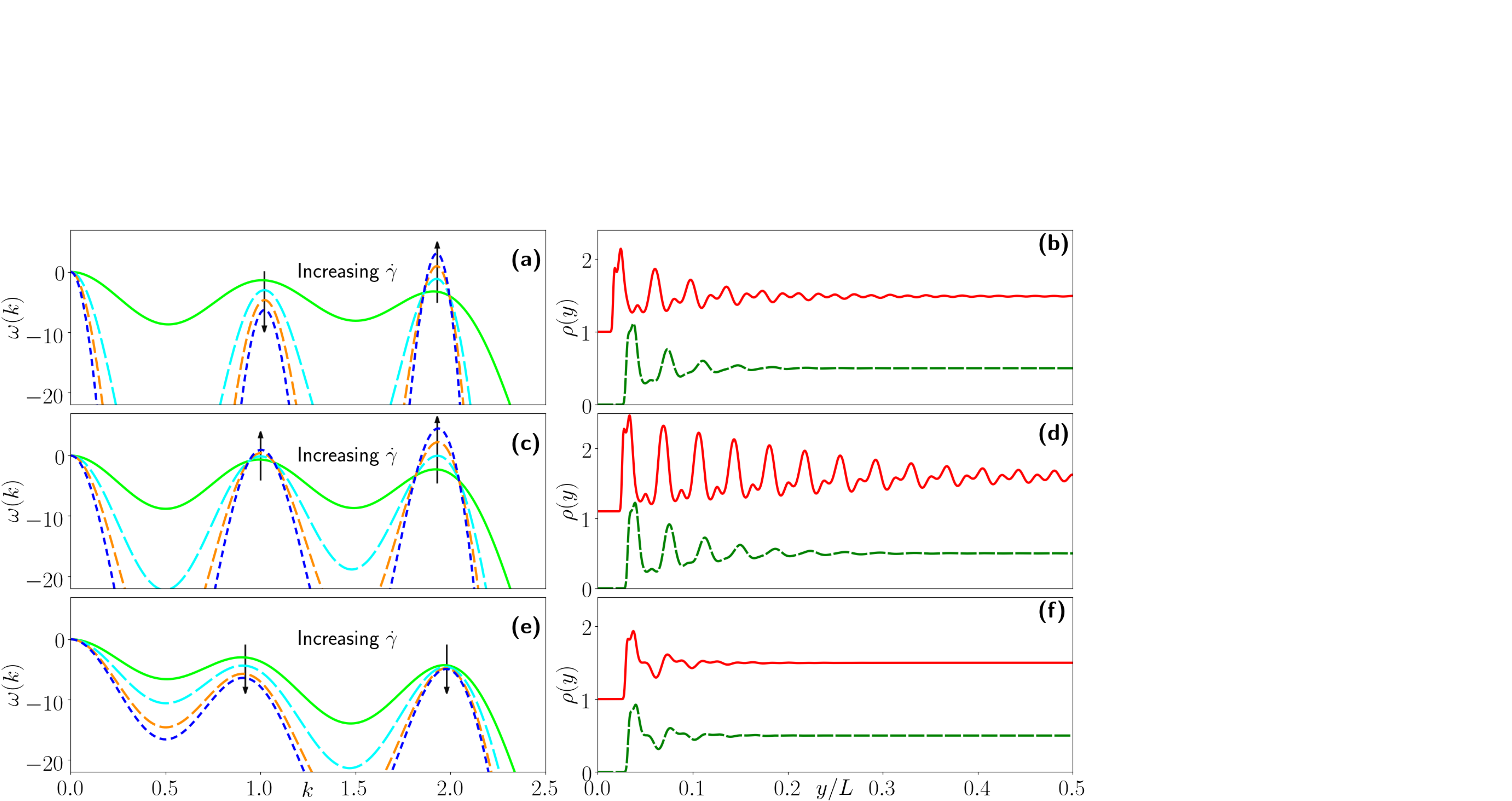}\caption{On the left we display the dispersion relation $\omega(k)$ for varying shear rate $\dot{\gamma}$. In each case the solid line is the unsheared $\dot{\gamma}=0$ case, followed by the long dashed, intermediate dashed, and short dashed lines, for increasing $\dot{\gamma}$,  respectively. On the right we display corresponding density profiles from DDFT, where the dashed line is for $\dot{\gamma}=0$ (unsheared equilibrium system) and the solid line is the steady state for particular $\dot{\gamma}>0$ (vertically shifted by 1, except for in (d), which is shifted by 1.1). Three cases are considered: \emph{Case 1} with parameters $\mathcal{C}_1$, displayed in (a) and (b); \emph{Case 2}, for $\mathcal{C}_2$, in (c) and (d); and \emph{Case 3} with parameters $\mathcal{C}_3$, in (e) and (f). In \emph{Case 1} the $\omega(k)$ displayed in (a) is for shear rates $\dot{\gamma}=0$, $1.5$, $3$ and $4.5$. As $\dot{\gamma}$ is increased the first peak at $k\approx 1$ decreases, while the second at $k\approx2$ increases. For strong enough shear one can eventually swap from one dominant length-scale to the other. In (b) the density profiles are for $\dot{\gamma}=0$ and $\dot{\gamma}=4.5>\dot{\gamma_c}$. In the sheared suspension the short (wavelength) length-scale becomes more prominent and the long length-scale less so. In (c) the results are for $\dot{\gamma}=0$, $0.5$, $1$ and $1.5$. As $\dot{\gamma}$ is increased, both peaks in $\omega(k)$ increase in value, leading eventually to the system becoming linearly unstable. In (d) the \emph{Case 2} density profiles are for $\dot{\gamma}=0$ and $\dot{\gamma}=1.5 >\dot{\gamma_c}$. In the sheared case, both length-scales are more prominent and have a slower decay away from the wall. Moreover, the short length-scale (almost not visible for $\dot{\gamma}=0$) becomes slightly dominant. For \emph{Case 3} in (e), the $\omega(k)$ displayed are for $\dot{\gamma}=0$, $0.2$, $0.4$ and $0.5$. As $\dot{\gamma}$ is increased, both peaks in $\omega(k)$ move down, making all density modulations even more strongly damped, so decreasing any structuring away from the wall. The corresponding density profiles in (f) are for $\dot{\gamma}=0$ and $0.5$.}\label{full_fig}
\end{figure*}

\emph{Case 1:} In order to achieve a switch between the two length-scales we set the pair potential parameters as $\mathcal{C}_1 \equiv \lbrace \sigma+0.006, c_0, c_2, c_4, c_6, c_8 \rbrace$. This choice has been made by studying the form of the first term in Eq.~(\ref{dispersion_relation_final}). Figure~\ref{full_fig}(a) shows the dispersion relation $\omega(k)$ for increasing shear. The solid (green) line is the static $\dot{\gamma}=0$ case. It has two peaks at $k_1$ and $k_2$, corresponding to the two intrinsic length-scales in the system. The least damped length-scale, and thus the highest peak, corresponds to $k=k_1=1$. In order to diminish this peak (i.e.\ make modes with this wavenumber more damped) and enhance the smaller length-scale (with wavenumber $k_2\approx 2$), the first term of Eq.~(\ref{dispersion_relation_final}) must be negative at $k_1$ and positive around $k_2$. For the pair potential parameters $\mathcal{C}_1$, on imposing an external shear on the system, the $k\approx 1$ peak in the dispersion relation decreases (see Fig.~\ref{full_fig}(a)). Simultaneously, the $k\approx 2$ peak increases and eventually the system becomes linearly unstable, when $\omega(k_2)>0$. This effect can be obtained for a range of values of $\delta\sigma$ or for other more complex variations of the set $\mathcal{C}_{R}$. However, here our aim is to use parameter sets close to the reference one.

In Fig.~\ref{full_fig}(b) we show the $\dot{\gamma}=0$ equilibrium density profile (dashed green) and the $\dot{\gamma}=4.5$ steady-state density profile (solid red), where the external potential parameter is $\beta V_0 =400$. The $\dot{\gamma}=0$ density profile contains oscillations (correlations) with both length scales, but because the large length scale $\sim2\pi/k_1$ is dominant, these are most easily visible in the profile, particularly further from the wall (larger $y$) -- c.f.\ Ref.~\cite{walters2018structural}. Away from the wall the density is uniform, because for this value the bulk fluid state is linearly stable. Increasing $\rho_b$ leads to the oscillations increasing in amplitude and decaying more slowly into the bulk, because one is approaching the linear stability threshold associated with freezing. However, under shear the small length-scale comes to dominate over the large length-scale, being increasingly prominent as $y$ increases. Moreover, only the oscillation with the small wavelength survive in the center of the system, setting the wavelength of the shear induced laning. The results in Figs.~\ref{full_fig}(a) and \ref{full_fig}(b) show that a system exhibiting one characteristic length-scale can be controllably tuned by means of shear to obtain a system exhibiting a different typical (lane) length-scale.

\emph{Case 2}: We now consider the reference set $\mathcal{C}_2=\mathcal{C}_{R}$. In this setup, the effect of the flow kernel is to increase both peaks in the dispersion relation, as shown in Fig.~\ref{full_fig}(c). In Fig.~\ref{full_fig}(d) we display corresponding density profiles, with $\beta V_0=500$, and $\dot{\gamma}=0$ and $\dot{\gamma}=1.5$. The $\dot{\gamma}=0$ equilibrium density profile shows the typical oscillations decaying towards the center of the system, with the wavelength being the large length-scale. The peak in the dispersion relation corresponding to this is the highest, being closer to zero. In the region close to the wall one can observe the contribution from the small length-scale density oscillation decaying relatively fast. However, as $\dot{\gamma}$ is increased, the small length-scale becomes more prominent and for a shear rate stronger than the critical shear $\dot{\gamma}_c$ (corresponding to the onset of the linear instability), we observe that both length-scales influence the structure deeper into the bulk. This example shows that it is possible to switch on and off multiple length-scales, which are not present (in the bulk) in the unperturbed situation.

\emph{Case 3}: The interesting feature of this case is that shear makes the system more stable than the equilibrium state, i.e.\ both of the peaks in $\omega$ are suppressed by shear. This phenomenon can be observed when the parameters $\mathcal{C}_3 \equiv \lbrace \sigma+0.006, c_0, c_2, c_4-0.0147, c_6, c_8 \rbrace$. As before, this set has been determined by studying the form of the first term in Eq.~(\ref{dispersion_relation_final}). Specifically, the effect of the shear is to  decrease the height of both peaks in the dispersion relation (Fig.~\ref{full_fig}(e)). In this situation the system is less susceptible to external perturbations and thus is able to absorb any modulation faster if subject to external shear. In Fig.~\ref{full_fig}(f) this mostly affects the rate of decay of the density oscillations into the bulk, rather than the initial amplitude, so the effect of the shear is not immediately apparent. This suppression of both peaks in $\omega(k)$ can be obtained with a range of values of ($\delta\sigma$, $\delta c_4$) or from other more complex variations of the set $\mathcal{C}_{R}$. However, our aim is to use sets close to the reference one.

To summarise, we have shown that it is possible to controllably tune the typical particle spacing in soft matter systems exhibiting multiple intrinsic length-scales, using shear as the control. For our soft particle model we have shown that depending on the form of the interactions (i.e.\ the chemical make-up of the molecules), it is possible to change the characteristic distance between lanes of particles. Our work shows that tiny changes in the chemistry of the particles, namely in the form of the pair interaction potential, can lead to extremely different macroscopic structures. We suggest that this non-linear behavior can be exploited to develop a class of active metadevices. The key conceptual step to perform this study is the identification of the shear kernel \citep{kruger2011controlling, brader_kruger_2011, scacchi_kruger_brader_2016, scacchi2017dynamical, scacchi2018flow, stopper2018nonequilibrium}. This only occurred recently  and it deserves much future study. 

{\it Acknowledgments}: AS is supported by the Swiss National Science Foundation under the grant number P2FRP2$\_$181453 and  AJA by the EPSRC under the grant EP/P015689/1.

\section*{Appendix}

\subsection*{Dispersion relation}\label{dispersion relation} 
Here we derive the dispersion relation $\omega(k)$ in Eq.~\eqref{dispersion_relation_final} for a sheared system. This comes out of performing a stability analysis on the generic DDFT equation with non-affine advection and follows the treatment in Ref.~\citep{scacchi2017dynamical}. $\omega(k)$ is the key quantity that allows to predict the characteristic length-scales in the system and how these are influenced by externally imposed shear. We start by assuming the one-body density takes the form of a constant plus a small perturbation,
\begin{equation}
\rho(\mathbf{r},t)=\rho_b+\delta\rho(\mathbf{r},t)=\rho_b+\sum_{\textbf{k}}\epsilon_{\textbf{k}} e^{i\textbf{k}\cdot \mathbf{r}+\omega t},\label{density_perturbation}
\end{equation}
where we have assumed a Fourier decomposition of the perturbation to write it as a sum over modes with wave vector $\textbf{k}$, corresponding amplitude $\epsilon_{\textbf{k}}$ and $\omega(k)$ is the growth/decay rate, referred to as the dispersion relation. For simplicity, we consider a 2-dimensional system in the $(x,y)$ plane. However, the generalization to 3-dimensions is straightforward. It is convenient to write the velocity field as:
\begin{equation}\label{velocity}
\textbf{v}(\mathbf{r})=v^{aff}(y)\textbf{e}_{x}+\textbf{v}^{fl}(\mathbf{r})=\dot{\gamma} y \textbf{e}_x + \textbf{v}^{fl}(\mathbf{r}),
\end{equation}
where a simple shear is assumed to be acting along the $x$-axis, where $\dot{\gamma}$ is the shear rate and $\textbf{v}^{fl}(\mathbf{r})$ is the non-affine term (also referred to as the `fluctuation' term). Using these definitions, and a functional Taylor expansion of $c^{(1)}(\mathbf{r})$, i.e.
\begin{equation}
c^{(1)}(\mathbf{r})=c^{(1)}[\rho_b]+\int d\mathbf{r}' \delta\rho(\mathbf{r}')c^{(2)}(\mid \mathbf{r}-\mathbf{r}'\mid)+\mathcal{O}(\delta\rho^2),
\end{equation}
where $c^{(2)}(r)=\delta c^{(1)}/\delta\rho\mid_{\rho_b}$ is the bulk fluid two-body direct correlation function, Eq.~\eqref{DDFT} reduces to:
\begin{equation}
\begin{split}
\omega\delta\rho(\mathbf{r},t)&
+\rho_b\nabla\cdot\textbf{v}(\mathbf{r})+\nabla\cdot\left[\delta\rho(\mathbf{r},t)\textbf{v}(\mathbf{r})\right]\\&
=-k^2 D\left[1-\rho_b\hat{c}(k)\right]\delta\rho(\mathbf{r},t)+\mathcal{O}(\delta \rho^2),
\end{split}
\end{equation}
where $D=k_\mathrm{B} T\Gamma$ is the diffusion coefficient and $\hat{c}(k)$ is the Fourier transform of $c^{(2)}(r)$. The right hand side of the last equation is a known result~\cite{marconi_tarazona_1, archer_evans, archer2012solidification, evans_79, andy_walters_thiele_knobloch}. The Taylor expansion of the free energy is an appropriate approximation, since we assume the density variations around the bulk value to be small, i.e.\ $\delta\rho\ll1$. We then recall that for an equilibrium fluid $\left[1-\rho_b\hat{c}(k)\right]=1/S(k)$, where $S(k)$ is the static structure factor~\cite{evans_92}. 
Using Eq.~(\ref{velocity}) and then dividing each term by $\delta\rho(\mathbf{r},t)$ we obtain
\begin{equation}
\begin{split}
\omega
=&-\frac{\rho_b}{\delta\rho(\mathbf{r},t)}\nabla\cdot\textbf{v}^{fl}(\mathbf{r})-ik_x v^{aff}_x(\mathbf{r})-ik_x v^{fl}_x(\mathbf{r})\\&
-ik_y v^{fl}_y(\mathbf{r})-\nabla\cdot\textbf{v}^{fl}(\mathbf{r})-\frac{k^2D}{S(k)}+\mathcal{O}(\delta \rho).
\end{split}
\end{equation}
Since we are interested in the growth/decay of perturbations perpendicular to the flow we consider the wave vectors $\textbf{k}=(0,k_y)$. We then have
\begin{equation}
\begin{split}
\omega
=&-\frac{\rho_b}{\delta\rho(\mathbf{r},t)}\nabla\cdot\textbf{v}^{fl}(\mathbf{r})-ik_y v^{fl}_y(\mathbf{r})-\nabla\cdot\textbf{v}^{fl}(\mathbf{r})\\&
-\frac{k_y^2D}{S(k_y)}+\mathcal{O}(\delta \rho).
\label{omega1}
\end{split}
\end{equation}
The non-affine velocity term has the general convolution form shown in Eq.\ (5). Replacing Eq.\ (\ref{density_perturbation}) in Eq.\ (5), and using $\mathbf{r}''=\mathbf{r}-\mathbf{r}'$ we find
\begin{equation}
\textbf{v}^{fl}(\mathbf{r})=\delta\rho(\mathbf{r},t)\int d\mathbf{r}''\left[\cos(\textbf{k}\cdot\mathbf{r}'')-i\sin(\textbf{k}\cdot\mathbf{r}'')\right]\boldsymbol{\kappa}(\mathbf{r}'').
\end{equation}
We recall that the $y$-component of the kernel function, $\kappa_y$, has to be an odd function, which implies that $\int d\mathbf{r}' \kappa_y(\mathbf{r}-\mathbf{r}')=0$ and also that the first of the above integrals is vanishing, so that we can simplify to obtain
\begin{equation}
\begin{split}
\textbf{v}^{fl}(\mathbf{r})&=-i\delta\rho(\mathbf{r},t)\int d\mathbf{r}'' \sin(\textbf{k}\cdot\mathbf{r}'')\boldsymbol{\kappa}(\mathbf{r}'')\\&=-i\delta\rho(\mathbf{r},t)\boldsymbol{\alpha}(\textbf{k}),\label{v_fl_integral}
\end{split}
\end{equation}
where $\boldsymbol{\alpha}(\textbf{k})=\int d\mathbf{r}\sin(\textbf{k}\cdot\mathbf{r})\boldsymbol{\kappa}(\mathbf{r})$. In Eq.~(\ref{omega1}) the second and the third terms in the right hand side are of order $\delta\rho$, which can be neglected, since we consider the limit of small perturbation $\delta\rho$, i.e.\ $\epsilon_{\textbf{k}}\rightarrow 0$. We are left with the equation
\begin{equation}
\omega(\textbf{k})=-\rho_b \textbf{k}\cdot\boldsymbol{\alpha}(\textbf{k})-\frac{k_y^2D}{S(k_y)}.
\end{equation} 
Since we have chosen to consider the wave vectors $\textbf{k}=(0,k_y)$ in order to study the instability along the $y$-axis, we can write
\begin{equation}
\omega(0, k_y)=-\rho_b k_y \alpha_y(k_y)-\frac{k_y^2D}{S(k_y)},
\label{dispersion_relation_final_appendix}
\end{equation}
where $\alpha_y(k_y)=\int d\mathbf{r} \sin(k_y y)\kappa_y(\mathbf{r})$.

\subsection*{Flow kernel for uniaxial shear}
When imposing a uniaxial shear, i.e.\ when $\textbf{v}^{aff}(\mathbf{r})=y\dot{\gamma}\textbf{e}_x$, one can reduce the  problem to an effective one-dimensional system, by assuming that the density varies only along the stability axis, i.e.\ perpendicular to the walls responsible for the shear. The fact that some of the integrals do not depend on $\rho(\mathbf{r})=\rho(y)$ means that the velocity fluctuation term can be written as
\begin{equation}
v^{fl}_y(y)=\dot{\gamma}\int_{-\infty}^{\infty} \rho(y-y') \mathcal{L}(y') dy',
\end{equation}
where the function $\mathcal{L}(y)$ can be obtained numerically (only once, independently from the  shear rate $\dot{\gamma}$ and the one-body density $\rho$), reducing the full two-dimensional or three-dimensional problem to an effective (much quicker) one-dimensional one. For details see the Appendix in \citep{scacchi2017dynamical}.



%

\end{document}